\documentclass[aps,prl,twocolumn,superscriptaddress,10pt]{revtex4-2}
\usepackage[usenames,dvipsnames]{xcolor}
\usepackage{graphicx}  
\usepackage{epstopdf}  
\usepackage{overpic}   
\usepackage{dcolumn}   
\usepackage{bm}        %
\usepackage{amssymb}   %
\usepackage{textcomp}
\usepackage{color}   
\usepackage{amsmath}

\begin{document}




\title{Convective Nature of the Stimulated Raman Side Scattering in Inertial Confinement Fusion}

\author{F.-X. Zhou}
\affiliation{Department of Modern Mechanics, University of Science and Technology of China, Hefei, Anhui 230026, China}

\author{C.-W. Lian}
\affiliation{Department of Modern Mechanics, University of Science and Technology of China, Hefei, Anhui 230026, China}

\author{R. Yan}
\email{ruiyan@ustc.edu.cn}
\affiliation{Department of Modern Mechanics, University of Science and Technology of China, Hefei, Anhui 230026, China}

\author{Y. Ji}
\email{yji@ustc.edu.cn}
\affiliation{Department of Modern Mechanics, University of Science and Technology of China, Hefei, Anhui 230026, China}

\author{J. Li}
\affiliation{Department of Plasma Physics and Fusion Engineering, University of Science and Technology of China, Hefei 230026, China}

\author{Q. Jia}
\affiliation{Department of Plasma Physics and Fusion Engineering, University of Science and Technology of China, Hefei 230026, China}

\author{J. Zheng}
\affiliation{Department of Plasma Physics and Fusion Engineering, University of Science and Technology of China, Hefei 230026, China}


\date{\today}
\begin{abstract}
{The absolute growth of Stimulated Raman side scattering (SRSS) predicted by previous theories appeared to be surprisingly absent in the recent ignition-scale direct-drive experiments with the absence attributed to different reasons.
We present evidence from simulations that the linear SRSS modes are naturally all convective (i.e., absolute SRSS does not exist at all) in an experimentally relevant regime where a finite-beam-width laser is incident into a non-uniform low-density plasma below a quarter of the critical density. 
The convective gain demonstrated by our newly proposed formula via numerical fitting monotonically increases with the beam width without saturation, which is significantly different from the prediction of previous convective SRSS theories.
}
\end{abstract}
  

\maketitle



As the ignition milestone of inertial confinement fusion (ICF) \cite{Zylstra2022, Abu-Shawareb2023, Betti2023} has been achieved at the National Ignition Facility (NIF), limiting
harmful effects of laser-plasma instabilities (LPIs) remains critical in the next phases of a fusion energy future. 
LPIs are fundamental limiters of fusion performance with the negative impacts including the power loss due to the lights scattered away through Stimulated Raman Scattering (SRS) or Stimulated Brillouin Scattering (SBS), and the premature fuel preheating due to the energetic (hot) electrons that are accelerated by electron-plasma waves (EPWs) driven by SRS or two-plasmon decay (TPD). SRS is a three-wave coupling process through which an incident laser beam decays into a scattered light and an EPW while satisfying the matching conditions \cite{Kruer2003,Michel2023} for both frequencies and wavevectors. SRS is recognized as the dominant preheating process over TPD in ignition-scale direct-drive experiments \cite{Rosenberg2018} where the lasers directly irradiate the target.

A specific SRS geometry in which the scattered light is stimulated nearly perpendicular to the electron density gradient is often referred to as Stimulated Raman Side Scattering (SRSS) and has attracted recurring research interest over the past few years \cite{Rosenberg2018, Xiao2018, Michel2019, Rosenberg2020, Short2020, Xiao2023} as SRSS has been identified as a significant source in the SRS spectra in the recent ignition-scale direct-drive experiments conducted at the leading ICF facilities \cite{Rosenberg2018, Glize2023}. It was predicted in theory \cite{Liu1974} as early as in the 1970s that SRSS is the only type of SRS to become absolutely unstable in the low-density regions substantially below the quarter critical density ($n_c/4$) of the incident laser in an inhomogeneous plasma, while all types of SRS are well known to be able to grow convectively in the low-density regions. 

Before reaching nonlinear saturation, absolute modes grow exponentially with time at a fixed spatial location, whereas convective modes have limited factors of amplification on the initial seeds. It was theoretically identified in Ref.~\cite{Mostrom1979} that SRSS consists of both absolute and convective modes.
The predicted absolute feature and the low absolute thresholds demonstrated by a series of theories \cite{Liu1974, Mostrom1979, Afeyan1985} had made SRSS of particular concern in ICF by the 1980s. However, the lack of experimental evidence \cite{Mostrom1977, Mostrom1979} for SRSS under conditions where the laser intensity was above the absolute thresholds in the early experiments attenuated the research interest until recently when significant SRSS signals were observed in modern experiments \cite{Rosenberg2018, Rosenberg2020, Glize2023, Zhao2024}. Short \cite{Short2020} then extended the SRSS theory to the multiple-beam regime and predicted similar absolute thresholds as in Ref.~\cite{Liu1974, Afeyan1985} in the single-beam regime. 
It was realized in Ref.~\cite{Mostrom1979} that a finite beam width would lead to weaker growth of the absolute modes compared to the plane-wave regime (i.e., with an infinitely wide laser) \cite{Liu1974, Afeyan1985, Short2020}.  
Although significant SRSS signals were observed in recent experiments, these signals were attributed to convective rather than absolute modes \cite{Michel2019,Rosenberg2020}.
When absolute SRSS would experimentally show up remains a confusing question and brings uncertainty in ICF designs.

The surprising absence of absolute SRSS in experiments was attributed to different reasons. Mostrom and Kaufman \cite{Mostrom1979} suggested that the finite beam width limits the density range over which the absolute SRSS modes can grow to dominance 
and anticipated the importance of the convective SRSS modes. Michel et al. \cite{Michel2019} suggested that damping of the EPWs pushes the absolute instability threshold to higher laser intensities at lower densities under ignition-scale conditions and proposed a convective-gain model in the strong-damping limit to explain the SRSS signals originated in the low density regions in the NIF experiments. 
The gain formula in Ref.~\cite{Michel2019} was also implemented in the ray-tracing analysis on the experiments in the SG-II Upgrade facility and showing consistency with the angular spread of SRSS signals \cite{Glize2023}. 
The convective theories \cite{Mostrom1979,Michel2019,Xiao2023} based on a three-wave coupling system all predicted a finite upper limit of the gain as long as the laser beam is sufficiently wide, 
because a seed scattered light of SRSS refracts as it propagates in an inhomogeneous plasma and eventually leaves its resonant density range.

In this Letter, we present evidence from numerical simulations that the linear SRSS modes are naturally all convective (i.e., the absolute modes do not exist at all) in an experimentally relevant finite-beam-width regime in low-density plasmas below 0.25 $n_c$, for the first time. A new convective gain formula is then proposed via fitting of the simulations which demonstrate significantly different dependence on the beam width from the previous convective SRSS theories. A newly developed fluid code \textit{FLAME-MD} \cite{Zhou2023} and  a full particle-in-cell (PIC) code \textit{OSIRIS} \cite{Fonseca2002} are used for the simulations.
\textit{FLAME-MD} solving the set of fluid-like LPI equations \cite{Hao2017,Zhou2023} has been applied to accurately capture the linear growth and identify the nature of LPI modes 
in different regimes \cite{Zhou2023,Ji2023}. In \textit{FLAME-MD} simulations, we are able to eliminate all nonlinear or kinetic effects and obtain purely linear behaviors of SRSS. 

The fluid simulations in this Letter are mostly performed in a finite-beam-width regime: the plasma has a uniform electron temperature ($T_e$) and a linear electron density ($n_e$) profile typically ranging from 0.1 to 0.15 $n_c$ as $n_e(z) = (0.1+0.25z/L)n_{c}$, where $L$ is the density scale length $n_e/({d n_e}/{d z})$ evaluated at $n_e = n_c /4$. A prescribed laser beam with the wavevector $\bm{k}_0$ and frequency $\omega_0$  propagates along the electron density gradient ($\nabla n_e$) with the electric field polarized along $x$. Ions are kept static in all fluid simulations to ensure that SRS is the only possible LPI in the low-density plasma. 
The laser intensity is prescribed with a transverse Gaussian profile having  the maximum intensity ($I_0$) on the axis, such that $I_0$ is invariant in time and has no pump depletion, ensuring that SRS always evolves in the linear regime in the fluid simulations. 

\begin{figure}[b]
	\begin{center}
		\includegraphics[width=3.375in]{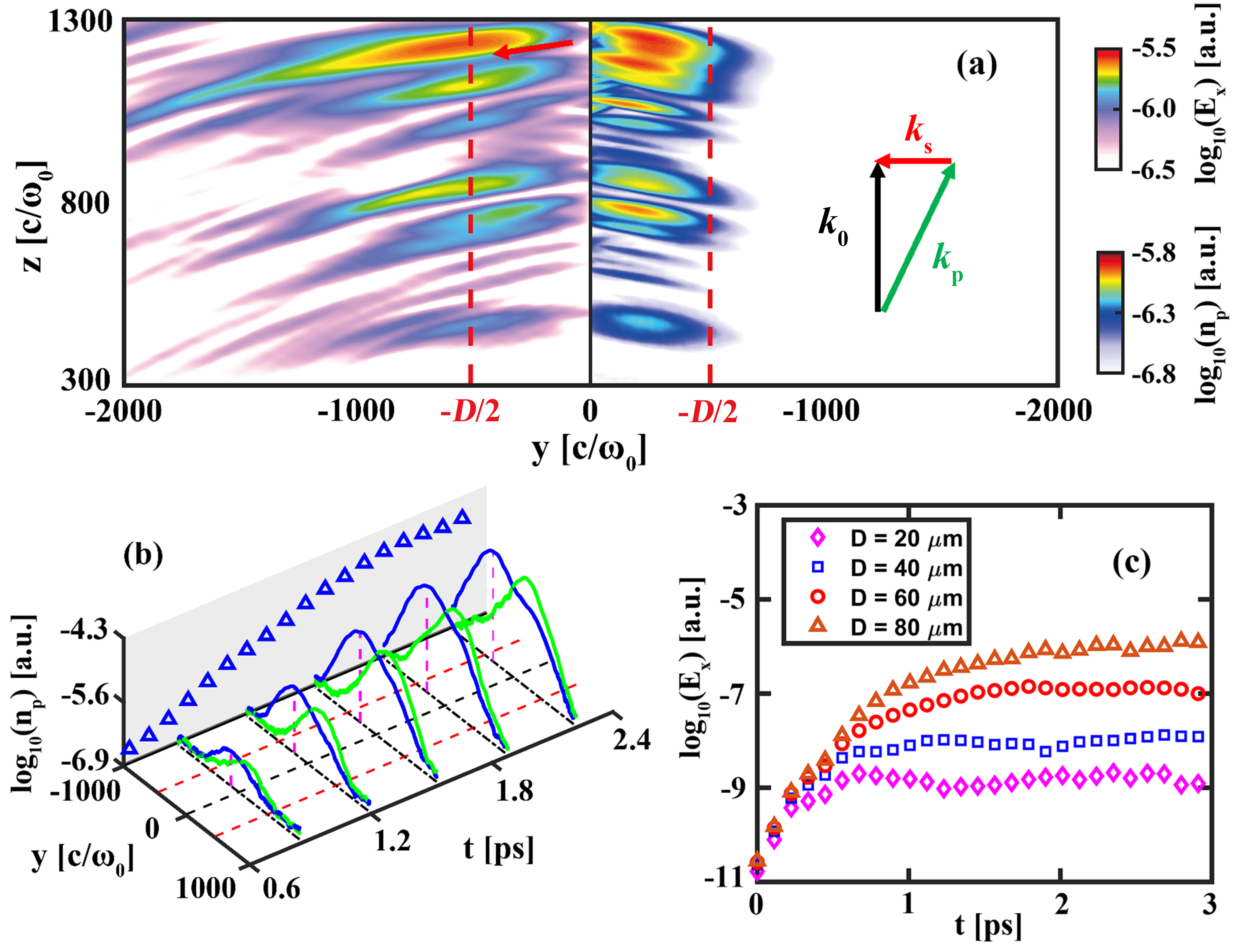}
\caption{ 
	(a) Envelope of the electric field of scattered lights (left half) and  envelope of the electron density perturbations (right half, mirrored) associated with the EPWs at $t$ = 1.1 ps in the NIF-fluid case.  The red dashed lines illustrate the incident laser's width with $D = 60\mu$m. The wavevector matching condition $\bm{k}_0 = \bm{k}_p + \bm{k}_s$ is also sketched, where the subscripts $0$, $p$, and $s$ represent the incident light, the EPW, and the scattered light, respectively. (b) Temporal and spatial evolution of the local EPWs' envelopes [blue (green) line: left (right)-shifting] in a narrow density region of $0.125 \sim 0.135$ $n_c$ in the NIF-fluid case. Blue triangles: the peak amplitudes of the left-shifting envelopes.  (c) Temporal evolution of the electric field amplitude $E_x$ of the side-scattered lights with $k_y=0.5 \omega_0/c$ in the fluid cases with the same other parameters ($\lambda_0$ = 1.054 $\mu$m, $I_0 = 3 \times 10^{15}$ W/cm$^2 $, $L=100$ $\mu$m, and $T_e= 2$ keV) but different $D$.}
\label{fig1}
	\end{center}
\end{figure}

We first show the fluid simulation (referred to as the NIF-fluid case) with the NIF-relevant laser-plasma parameters  $I_0 = 1.3 \times 10^{15}$ W/cm$^2 $, $\lambda_0$ = 0.351 $\mu$m,  $L=525$ $\mu$m, and $T_e= 4.5 $ keV, where $\lambda_0$ is the laser wavelength in vacuum. This set of parameters is relevant to that reported in Ref.~\onlinecite{Rosenberg2018}. The simulation box is 268 $\mu$m (y) $\times$ 83 $\mu$m (z) with a grid of 24000 $\times$ 7500. 
The laser beam has a diameter $D = 60\mu$m as illustrated by the dashed lines in Fig.~\ref{fig1}(a) where the laser intensity drops to $I_0/e^2$ at $y= \pm D/2$. Perfectly matched layers (PML) \cite{Berenger1994} are applied as the open boundary conditions for the scattered lights on all boundaries to ensure the scattered lights freely leave the simulation domain without recirculation or reflection. Neither collisional nor Landau damping is included in this simulation to exclude possible inhibition of absolute SRSS modes due to damping \cite{Michel2019}.

Significant growth of SRSS is found in the NIF-fluid case, as shown in Fig.~\ref{fig1}(a). The envelopes of the electric field of the scattered lights and the electron density perturbation ($n_p$) associated with the EPWs due to SRSS in the early stage ($t = 1.1 $ps) are illustrated in the left and right halves of Fig.~\ref{fig1}(a). The simulation is essentially symmetric with respect to the axis of the laser thus only  half of the simulation domain is drawn with $n_p$ mirrored for side-by-side illustration. The envelopes are obtained by picking the magnitude of the matching-condition-predicted most resonant SRSS modes which are also found to be the dominant modes in the Fourier spectra in sweeping windows.
It is shown in Fig.~\ref{fig1}(a) that the side-scattered lights are propagating sidewards meanwhile refracting toward lower density (demonstrated by the red arrow) due to the density inhomogeneity along $z$. The scattered lights reach their maximum amplitudes near the left edge ($y=-D/2$) of the laser at this moment, evidencing that the scattered lights could no longer grow once they propagate out of the incident laser beam.
The EPWs satisfying the matching conditions sketched in Fig.~\ref{fig1}(a) are located on the propagation paths of the scattered lights, evidencing that the EPWs and scattered lights are the paired daughter waves of SRSS. 
The maximum amplitudes of the EPWs are reached closer to the laser axis, indicating that the EPWs are driven by the incident laser with a Gaussian profile and the scattered lights.  Furthermore, the growth of the backward scattered lights in the simulation is found minimal, consistent with the small Rosenbluth gain \cite{Rosenbluth1972} of the Stimulated Raman Back Scattering for this set of parameters.

The growth of the SRSS modes is identified as convective. The temporal and spatial evolution of the local EPWs satisfying the matching condition in a narrow density region of $0.125 \sim 0.135  n_c$ is shown in Fig.~\ref{fig1}(b). 
The envelopes of $n_p$ of the EPW modes paired with the left(right)-propagating scattered lights are plotted as the blue (green) lines in Fig.~\ref{fig1}(b), which demonstrates typical convective behaviors: firstly, the locations of the maximum amplitudes of EPWs are not fixed in $y$ but shift away from the laser axis; secondly, the maximum amplitudes of EPWs do not keep growing but saturate to a finite value as shown by the blue triangles. Moreover, although $I_0 = 1.3 \times 10^{15}$ W/cm$^2 $ is much higher than the threshold intensity ($I_{th} = 2.5 \times 10^{13}$ W/cm$^2 $) predicted by Ref.~\cite{Short2020}, no absolute growth is seen in this case.

\begin{figure}[b]
	\begin{center}
		\includegraphics[width=3.375in]{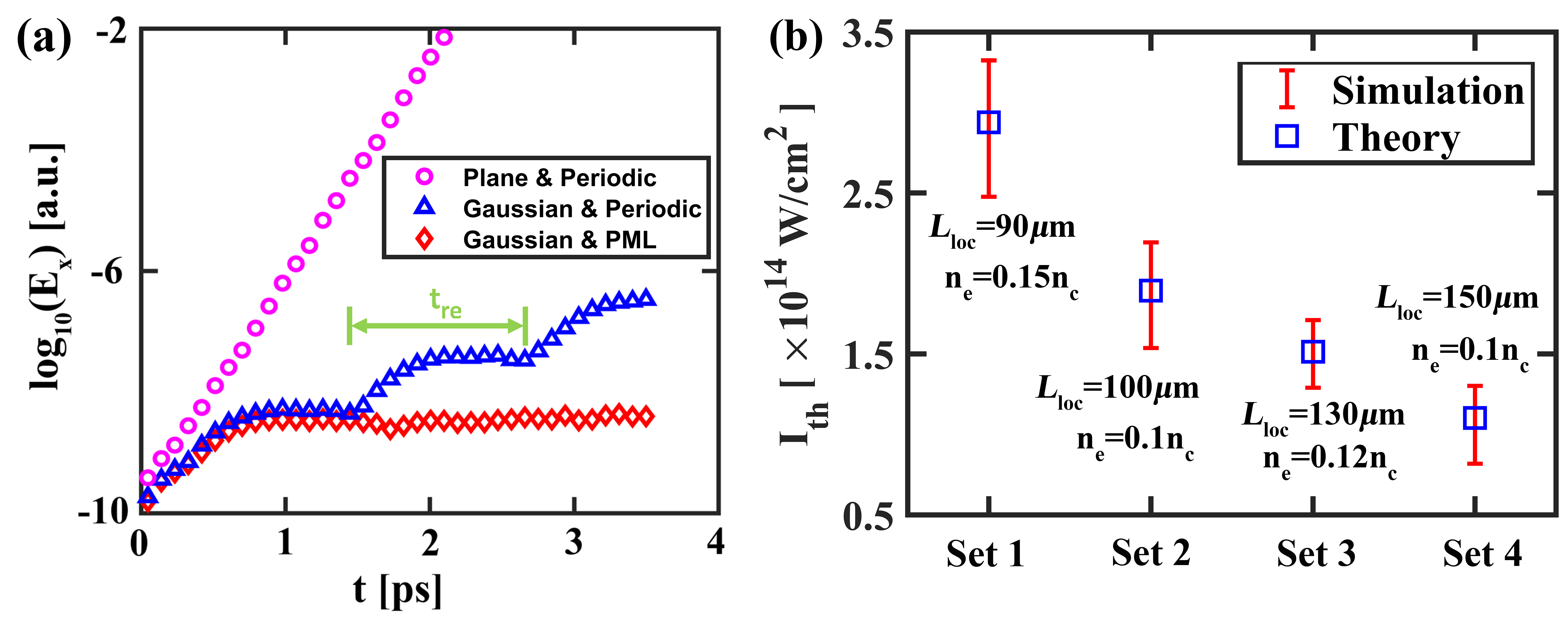}
		\caption{
	(a) Temporal evolution of $E_x$ for the side-scattered lights with $k_y=0.5 \omega_0/c$ in three fluid simulations with the same OMEGA-relevant parameters ($\lambda_0$ = 0.351 $\mu$m, $I_0 = 1.5 \times 10^{15}$ W/cm$^2 $, $L=150$ $\mu$m, and $T_e= 2$ keV) but different laser profiles and boundary conditions.
	The Gaussian laser profile has  $D=100$ $\mu$m. 
	(b) Absolute SRSS threshold intensities ($I_{th}$) at four sets of local electron densities and local density scale lengths ($L_{loc}$) in the plane-wave regime. Red bar: the range determined by the fluid simulations; blue square: $I_{th}$ predicted by Ref.~\cite{Short2020}. 
	In all sets $T_e = 2$ keV and $\lambda_0$ = 0.351 $\mu$m.}
\label{fig2}
	\end{center}
\end{figure}

Other than the NIF-fluid case, the fluid simulations performed for another set of parameters ($\lambda_0$ = 1.054 $\mu$m, $I_0 = 3 \times 10^{15}$ W/cm$^2 $, $L=100$ $\mu$m, and $T_e= 2$ keV) demonstrate no absolute growth either, despite the amplitude of the incident laser's electric field is extraordinarily large and favoring LPI growths ($I_0 \gg I_{th} = 2.7 \times 10^{14}$ W/cm$^2$ predicted by Ref.~\cite{Short2020}). The boundary conditions are the same as in the NIF-fluid case, without any damping included. All of these simulations with different $D$ demonstrate typical convective behaviors, as shown in Fig.~\ref{fig1}(c). Fig.~\ref{fig1}(c) plots the temporal evolution of the electric field amplitude ($E_x$) of the side-scattered lights with $k_y=0.5 \omega_0/c$, which is the resonant mode propagating perpendicular to the density gradient around 0.13$n_c$ determined by the matching conditions for SRSS. 
It is shown in Fig.~\ref{fig1}(c) that $E_x$ first grows exponentially, then saturates and reaches a convective saturation.
Moreover, the saturation level is found larger with larger $D$.

The theoretically predicted \cite{Mostrom1979, Liu1974, Afeyan1985, Short2020} absolute SRSS modes which are absent in all of our simulations above are found to show up if the simulation is configured using a plane-wave incident laser with periodic transverse boundary conditions, as shown by the circles in Fig.~\ref{fig2}(a). Fig.~\ref{fig2}(a) plots the temporal evolution of $E_x$ for the mode with $k_y=0.5 \omega_0/c$ under different laser profiles and boundary conditions in a set of fluid simulations with OMEGA\cite{Boehly1997, Craxton2015}-relevant parameters ($\lambda_0$ = 0.351 $\mu$m, $I_0 = 1.5 \times 10^{15}$ W/cm$^2 $, $L=150$ $\mu$m, and $T_e= 2$ keV). Continuous exponential growth without saturation (i.e., the absolute mode) is shown by the circles from the plane-wave case which satisfies the plane-wave assumption that the theories were based on. Absolute growth appears in the plane-wave case as predicted by Ref.~\cite{Short2020} since  $I_0 > I_{th}= 1.3 \times 10^{14}$ W/cm$^2$ with the OMEGA parameters. This is also consistent with earlier simulations \cite{Klein1973,Biskamp1975}, which demonstrated the existence of absolute SRSS in the plane-wave regime.

It is further shown that the thresholds for absolute SRSS in the simulations agree well with the theory \cite{Short2020} in Fig.~\ref{fig2}(b). 
To find the upper and lower bounds of $I_{th}$, we have performed four sets of fluid simulations with different central densities ($n_{mid}$) in the simulations and local density scale lengths ($L_{loc}$) [i.e.,  $L_{loc} = n_e/({d n_e}/{d z})$ evaluated at $n_e = n_{mid}$] which are needed by the threshold formula of Ref.~\cite{Short2020}.
It is found that $I_{th}$ given by the Ref.~\cite{Short2020} (blue squares) lies within the range determined by the simulations (red bars) in all cases, evidencing that the absolute growths exist in the plane-wave regime once $I_0>I_{th}$ as theoretically predicted. 

Similar to the convective behaviors shown in Fig.~\ref{fig1}(c), the absolute growth vanishes when the plane-wave laser is replaced by a Gaussian laser ($D = 100 \mu$m) and the PML boundary conditions are applied, see the diamonds in Fig.~\ref{fig2}(a). Typical convective behaviors are demonstrated as $E_x$ eventually reaches a saturation later on. We then switch the transverse boundary condition to periodic and obtain the temporal evolution of $E_x$ plotted with the blue triangles in Fig.~\ref{fig2}(a). The periodic boundaries allow the scattered lights of SRSS to re-enter the computational domain and to be convectively amplified again and again. The blue triangles demonstrate that $E_x$ in the periodic case has similar growth and saturation as in the PML case in the early stage ($t<1.5$ ps) but repetitively undergoes growth and saturation later on. The period between the onsets of two successive growths $t_{re} = 1.2$ ps is found approximately equal to the so-called recirculation time, i.e., the time needed for the side-scattered lights traveling across the width of the simulation box. The re-growths are therefore attributed to the multiple recirculation and reamplification of the side-scattered lights, which leads to temporally unbounded (i.e., absolute-like) growths in the finite-beam-width simulations with periodic transverse boundary conditions. However, neither this absolute-like growth artificially caused by the periodic boundary condition nor the absolute mode in the plane-wave regime is expected to be applicable in experimental conditions.

\begin{figure}[th]
	\begin{center}
		\includegraphics[width=3.375in]{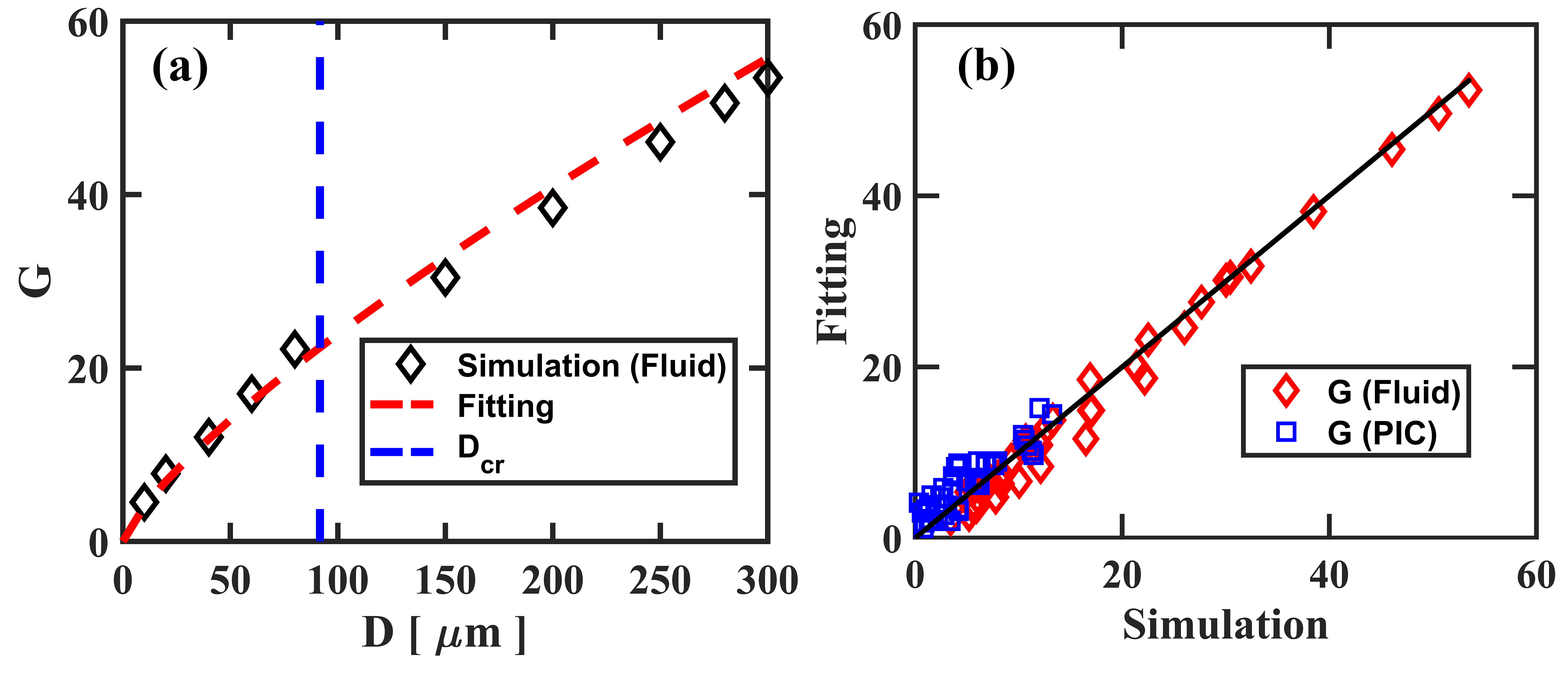}
		\caption{
	(a) Diamonds: $G$ of the side-scattered light with $k_y = 0.5 \omega_0/c$ versus $D$  in a set of fluid simulations with the parameters $\lambda_0$ = 1.054 $\mu$m, $I_0 = 3 \times 10^{15}$ W/cm$^2 $, $L=100$ $\mu$m, and $T_e= 2$ keV. Red dashed: $G$ given by Eq. (\ref{eq:G}).  The critical diameter $D_{cr}$ \cite{Xiao2018} is marked by the blue dashed line. (b) Fitted $G$ by Eq.(\ref{eq:G}) versus simulation data. 
	The simulation parameters covers: $5 \times 10^{14}$ W/cm$^2$  $\leq I_0 \leq $  $3 \times 10^{15}$ W/cm$^2$, 10 $\mu$m  $\leq D \leq $  300 $\mu$m, 50 $\mu$m  $\leq L \leq $  525 $\mu$m, 0.08 $n_c$ $\leq n_e \leq $  0.18 $n_c$, 1 keV $\leq T_e \leq $  4.5 keV, 0 $\leq \nu \leq $  $1 \times 10^{-3}$  $\omega_0$, and $\lambda_0$ = 0.351 or 1.054 $\mu$m.}
\label{fig3}
	\end{center}
\end{figure}

As the SRSS modes have been shown to be all convective in the finite-beam-width regime with open boundaries eliminating the recirculation of side-scattered lights, a plane-wave limit on the convective SRSS gain ($G$) arises such that $G$ is expected to approach infinity as $D\rightarrow \infty$, in order to be consistent with the plane-wave regime where the absolute growths indeed exist as predicted by the theories \cite{Liu1974, Mostrom1979,Afeyan1985,Short2020} and verified by our plane-wave simulations. 
However, all of the convective SRSS theories \cite{Mostrom1979, Michel2019, Xiao2018} predicted that $G$ reaches a finite value as long as $D$ exceeds a resonance width, which is  unable to recover the plane-wave limit.
We then perform fluid simulations with a range of $D$ to address this inconsistency and have found no saturation of $G$ even when $D$ is rather large, see Fig.~\ref{fig3}(a).  

Figure \ref{fig3}(a) plots $G\equiv 2 \ln (E_{sat}/E_{noise})$ versus $D$ with the same other simulation parameters detailed in the caption. Here $E_{sat}$ and $E_{noise} $ are the saturated level and the initialized noise level of $E_x$ of the matching-condition-predicted most resonant side-scattered lights at a specific $n_e$, respectively. The product of the full amplification time and the $y$ component of the group velocity of the side-scattered light  was considered as the critical diameter ($D_{cr}$) required for SRSS to achieve sufficient convective amplification thus a finite $G$ was expected when $D \ge D_{cr}$ \cite{Mostrom1979, Xiao2018}. However, Fig.~\ref{fig3}(a) shows that $G$ still increases with $D$ even if $D$ is a few times larger than $D_{cr}$ (marked by the blue dashed) rather than saturating at a finite value as expected by the convective theories \cite{Mostrom1979,Xiao2018,Michel2019,Xiao2023}. This trend demonstrated in our simulations is actually consistent with the plane-wave limit, suggesting a new form of convective gain formula.

A convective gain formula for SRSS is obtained by fitting more than 90 fluid and PIC simulations over a broad range of parameters, as shown in Fig.~\ref{fig3}(b). The PIC simulations which include self-consistent collisional and Landau damping are incorporated  not only to verify the fluid simulation results but also to help find the dependence of $G$ on damping, while the fluid simulations are phenomenologically implemented with effective damping. The PIC simulations have virtually the same grid resolutions and boundary conditions as the fluid simulations, and use 100 particles per cell.
The PIC simulation parameters are configured in lower-$G$ regimes with immobile ions to minimize possible nonlinear saturation mechanisms while the fluid simulations are free of nonlinearities such that each mode is naturally saturated by its linear convective amplification limit no matter how large $G$ is.
A scaling formula for $G$ 
is obtained by fitting the data of all simulations as
\begin{equation}
	\label{eq:G}	
	G=\frac{13.5  (\frac{v_0}{c})^2
 (\frac{D}{\lambda_0})^{0.77} (\frac{L}{\lambda_0})^{0.78} (\frac{n_e}{n_c})^{0.8}} { (\frac{v_e}{c})^{0.42} ( 1+860 \frac{\nu}{\omega_0}  )^{0.5}},
\end{equation}
where $v_0$ is the electron oscillatory velocity in the incident laser, ${v}_e$ is the electron thermal velocity, and $\nu$ is the effective damping rate including both collisional and Landau damping on the EPW. The density range covered by the simulations is 0.08 $n_c$ $\le n_e \le$ 0.18 $n_c$. Good fitting quality is reached as demonstrated by Fig.~\ref{fig3}(a) and (b). The dependence $G \propto I_0 (\propto v_0^2)$ is consistent with the experimental diagnostics \cite{Rosenberg2020} that SRS at lower densities has a near-exponential dependence on laser intensity. Moreover,
Eq. (\ref{eq:G}) demonstrates $G\rightarrow \infty$ as $D \rightarrow \infty$, which is substantially different from previous theories \cite{Mostrom1979,Michel2019,Xiao2018} and consistent with the plane-wave limit. The difference in the plane-wave limit implies that SRSS may not be sufficiently well modeled with a simple three-wave coupling system as the collective effects may contribute to the convective amplification, which requires further study in the future. 

While the absolute SRSS in the low-density plasma should not be a concern in the finite-beam-width regime any more, the gain of convective SRSS is expected by Eq. (\ref{eq:G}) to be significant in large-$L$ ignition-scale experiments, consistent with the experimental observations \cite{Rosenberg2018, Michel2019, Glize2023, Zhao2024}. The findings have significant implications for LPI assessment in ignition designs.

This research was supported by the National Natural Science Foundation of China (NSFC) under Grant Nos. 12388101, 12375242, 12375243, and U2430207, by the Strategic Priority Research Program of Chinese Academy of Sciences under Grant Nos. XDA25050400, XDA0380601, XDA25010200, and XDA25010100. The numerical calculations in this paper have been done on the supercomputing system in the Supercomputing Center of University of Science and Technology of China. We thank the UCLA-IST OSIRIS Consortium for the use of OSIRIS.

F.-X. Zhou and C.-W. Lian contributed equally to this work.

\bibliographystyle{apsrev}
\bibliography{SRSS-arxiv}

\newpage

\end{document}